\documentclass[11pt,conference]{IEEEtran}
\pdfoutput=1
\usepackage{color,graphicx}
\usepackage{amsmath}
\usepackage{enumerate}
\usepackage{syntonly}
\usepackage{multicol}
\usepackage{array}
\usepackage[tight,footnotesize]{subfigure}
\usepackage{url}
\usepackage{verbatim}

\makeatletter

\newcommand{\Rmnum}[1]{\expandafter\@slowromancap\romannumeral #1@}
\makeatother
\begin{document}
\title{Exploiting Semiconductor Properties for Hardware Trojans\\}
\author{\IEEEauthorblockN{Y. Shiyanovskii, F. Wolff, C. Papachristou}
\IEEEauthorblockA{Case Western Reserve University\\
Cleveland, Ohio 44106, USA\\
\{yxs32, fxw12, cap2\}@case.edu}
\and
\IEEEauthorblockN{D. Weyer, W. Clay}
\IEEEauthorblockA{Rockwell Automation\\
\{djweyer,sclay\}@ra.rockwell.com}}
\maketitle

\textbf{Abstract}. This paper discusses the possible introduction of hidden reliability defects during CMOS foundry fabrication processes that may lead to accelerated wearout of the devices. These hidden defects or hardware Trojans can be created by deviation from foundry design rules and processing parameters. The Trojans are produced by exploiting time-based wearing mechanisms (HCI, NBTI, TDDB and EM) and/or condition-based triggers (ESD, Latchup and Softerror). This class of latent damage is difficult to test due to its gradual degradation nature. The paper describes life-time expectancy results for various Trojan induced scenarios. Semiconductor properties, processing and design parameters critical for device reliability and Trojan creation are discussed. 

\begin{figure*}[htp]
\centering{
\includegraphics[scale=.55]{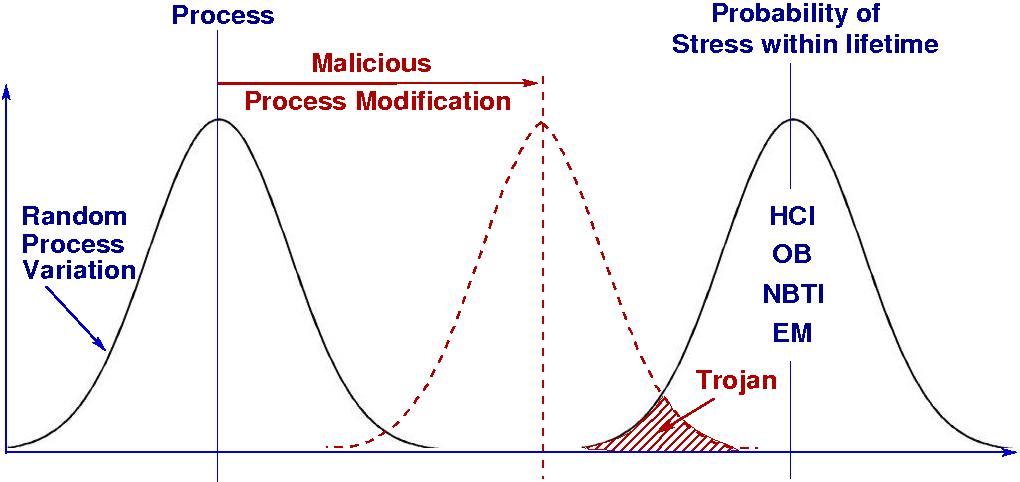}
}
\caption{Shifting the Process Variation to induce a Hardware Trojan based on Stress}
\label{f1}
\end{figure*}

\section{Introduction}
Due to the high capital costs of building and maintaining fabrication
facilities, the number of fabs is shrinking. 
Hence, there has been a major shift in control
over the fabrication process for integrated circuits (IC).
More and more vendors outsource the fabrication process to
off shore fabrication facilities \cite{TrojanPROB08A}.
Using such facilities makes the integrated circuits vulnerable
to malicious alterations.  These alterations are more commonly
known as hardware trojans and are usually created by insertion
of additional logic circuitry \cite{TrojanPROB08A}.
The intent of these trojans ranges from functional changes
of the circuit to a complete system failure.
Testing for Trojans at the manufacturing level have been
investigated using power analysis techniques \cite{HOST08g, HOST08a, HOST08e},
delay testing \cite{HOST08b}, and using test vectors \cite{HOST08d}.

In this paper we describe a new type of Trojan that  can be induced 
by intentional modification of fabrication process to accelerate
wearing processes in CMOS devices.  These process modifications
can keep the initial performance parameters of the integrated 
circuit within the accepted variation.   Such Trojans can exploit the following wearing processes:  Hot Carrier Injection (HCI), Oxide Breakdown (OB), Negative Bias 
 Temperature Instability (NBTI),  Electron Migration (EM).

IC manufactures optimize their processes to ensure the effects of the time based wear out mechanisms to guarantee the lifetime of a device for $10$ plus years over a specific temperature range i.e. $0$ to $70\,^{\circ}{\rm C}$ for commercial devices. IC manufactureÕs bound their process control limits to ensure production devices will not be affected by the time based wear out mechanisms for the guaranteed lifetime of the device. Figure \ref{f1} shows the normal distribution of the manufacturing process with the probability of a time based defect occurring within the guaranteed device lifetime.  

Production tests are not performed for these effects as they are time based, and the testing costs would be prohibitive. The production tests are optimized to find defects using the lowest cost equipment in minimal test time, and an acceptable test coverage to guarantee the devices are functional. The time based effects can be accelerated, but the test times are in days, which is not practical for production.  

A process engineer could maliciously modify certain steps in the process that would cause some or all devices on a wafer to wear out in months to few years, thus creating a reliability based Trojan. The paper will expound on process modifications that could be maliciously used to create reliability based Trojans. Figure \ref{f1} illustrates how the process engineer could modify a step in the manufacturing process to increase the probability that devices will be produced with a time based wear out mechanism. The parameters that affect these mechanisms are exponential in nature, so only small changes are needed in process steps. The overlap area in Figure \ref{f1} illustrates that a certain percentage of the devices on the wafer will be infected with these reliability Trojans.   
 
 \section{Wearing Mechanism of CMOS Devices}
In this section, we discuss the following wearing processes in CMOS devices. Hot Carrier Injection (HCI),
 Oxide Breakdown (OB), Negative Bias 
 Temperature Instability (NBTI),  Electron Migration (EM) We explore different 
degradation models and lifetime models for each process to identify critical dependencies that
can be exploited via a Trojan device by process characterization manipulation. 

\subsection{Hot Carrier Injection:}

The term hot carrier injection describes electrons (holes) 
that have accumulated sufficient kinetic energy
to overcome potential barrier and be injected into the gate oxide.
Such accumulation becomes more prominent in high electric field 
for electrons that have avoided subsequent scatterings with the lattice atoms.  
The carriers must overcome the $Si-SiO_2$ energy barrier of about $3.7eV$ for electrons and $4.6eV$ for holes.  For NMOS, hot electrons are produced and for PMOS holes are produced.  
  
Injection of hot carriers can result in the following: generation of new
 traps at or near the $Si-SiO_2$ interface or generation of new traps
  in the oxide itself.  The traps located at $Si-SiO_2$ interface affect the
 transconductance, $g_m$,  and leakage current of the device.  The traps that are located in the gate oxide increase the threshold voltage, $V_{th}$.  The carriers can also increase the substrate current, $I_{sub}$.
Thus, the HCI degradation can be monitored through shifts in the threshold voltage or transconductance and drain current. 

There are different models for predicting the lifetime of the device based on the rate of HCI effect.  One of such models uses a relationship between the shifts in gate current, $I_{gate}$, and the HCI degradation, \cite{ICCD1}. The model assumes that the rate of HCI damage to the CMOS device, $\Delta$, is proportional to the gate current, $I_{gate}$, and can be expressed by the following equation:

\begin{equation}
\label{e1}
\Delta / dt \sim I_{gate} =\frac {A(\Delta) } {W I_{drain}} * (\frac{I_{sub}}{I_{drain}})^m, 
\end{equation}
 where $W$ is the width of the CMOS device, \cite{ICCD4}, \cite{ICCD1}.  Abbreviating the following, 
 $B = A(\Delta) / W$ and merging that Median Time To Failure depends on the reciprocal of 
 $d\Delta / dt $, the failure rate is found from 
 
\begin{equation}
\label{e2}
 \lambda = BI_{drain} ( \frac{I_{sub}}{I_{drain}})^m
\end{equation}

The following equation assumes static voltages and currents, in order to determine dynamic
degradation, $\lambda$ must be integrated over a full cycle time, \cite{ICCD1}  Such model also does not include
the effect of operational temperature on the lifetime of the device.  However, the temperature does influence the rate of the HCI degradation and must be modeled accordingly.  In order to include temperature in the MTTF, the model must be based on the changes in substrate current.  The MTTF equation is the following:

\begin{equation}
\label{e3}
MTTF_{HCI} = B(I_{sub}) - N \exp(\frac{E_a}{kT}),
\end{equation}

where $B$ is a scale factor which is a function of doping profiles, sidewall spacings and dimensions that are specific to the manufacturing process characterization ($I_{sub}$) is the substrate current, $N$ ranges from 2 to 4,$k$ for the Boltzmann's constant, $T$ is temperature,  and $E_a$ is the activation energy in the range of $-0.1eV to -0.2eV$. 

There is another way to model HCI degradation, through a power law dependence on the stress time, $t$.  The HCI effect can be seen in shifts for the threshold voltage $V_{th}$ due to increased trap generation in the oxide and the interface.  Wang \cite{ICCD3} proposed the following model for the rate of change for the threshold voltage $V_{th}$ for NMOS devices:

 \begin{equation}
 \label{e4}
\Delta V_{th} \sim \sqrt{Q_i}\exp(\frac{E_{ox}}{E_o}) \exp (- \frac{\phi_{it}}{q\lambda E_m}) t^{n^{'}},
\end{equation}

where $E_{ox}$ is the oxide electric field, $E_m$ is the lateral electric field, $Q_i$ is the inversion charge, $\phi_{it}$ is the trap generation energy, $\lambda$ is the hot electron mean free path and $E_o$ is the process-dependent factor.  The model does not include the temperature ($T$) factor or the width ($W$) factor as seen in the $Equation (\ref{e3})$ and $Equation (\ref{e1})$ respectively.

 \subsection{Oxide/Dielectric Breakdown (OB):}
 
 In CMOS devices the strong electric fields across the gate oxide
  can cause some material damage that will lead to formation of a conductive path. The conductive path within the dielectric will create a short between the substrate and the gate, thus nullifying the isolating properties of the gate oxide and resulting in complete oxide failure.  The oxide breakdown 
  is a rather local phenomenon in which holes and bulk electron trapping damage the oxide material.  There are two types of oxide breakdown: early life oxide breakdown and time-dependent 
  dielectric breakdown.
  
The early life oxide breakdown is created through impurities and weak chains within the dielectric layer which is a direct result of the manufacturing process.  Any heavy alteration within the manufacturing process that will greatly increase trap generation within the oxide will lead to early oxide breakdown.  Increase in trap generation can be caused by the following factors: presence of mobile sodium ($Na$) ions in the oxide,  impurities trapped on the silicon interface during the oxidation, and crystalline 
  defects in the silicon.  
  Time-dependent dielectric breakdown (TDDB) consists of two phases: a build-up phase and a runaway phase,\cite{ICCD1}. During the build-up phase,  bulk electrons and holes gets randomly trapped through out the oxide.  The number of traps increase with time, forming a high electric field which depends on the voltage and oxide thickness\cite{ICCD5}.  
  
  \begin{equation}
  E_{ox} =\frac {V} { d_{ox}}
  \end{equation} 
  
  Once the trap density reached the critical trap density the runaway phase begins.  During the runaway phase, the  total electric field produced by the trapped electrons and holes exceeds the dielectric breakdown threshold,\cite{ICCD5}.  Once the electric field is strong enough, a conducting path is formed within the oxide and current begins to flow between the substrate and the gate.  Due to this behavior the time to total device failure due to oxide breakdown can be modeled using Weibull probability distribution. 
    
The rate of trap generation is the key component in determining the rate of oxide degradation and breakdown.  There are three general models for trap generation in the oxide: Anode Hole Injection (AHI) model, Thermo-Chemical (TC) model and Anode Hydrogen Release (AHR) model, \cite{ICCD1}  These models are contradictory in their explanation of the trap generation and the acceleration in the law of time-to-breakdown -  $t_{BD}$.  However two of the models, AHI (E model)  and TC (1/E model) can be summarized as follows: 

\begin{equation}
t_{BD} = \tau_{0} * \exp(-n * \gamma * E_{ox}^n)
 \end{equation}

where $\tau_{0}$ and $\gamma$ are constants and $n$ is either 1 for ($E$ model) or -1 for ($\frac{1}{E}$ model).  Both of these models are designed primarily for thick gate oxides. 

For ultra-thin oxides the primary driver of the breakdown process is gate voltage and at higher temperatures the process is accelerated even more.  To account for these factors, a new relationship has been proposed for the Median Time To Failure (MTTF) for the oxide breakdown:  

\begin{equation}
MTTF_{OB} = T_{BD0}(V) \exp(\frac{a(V)}{T} + \frac{b(V)}{T^{2}}),
\end{equation}

where $T_{BD0}$, $a$ and $b$ are voltage dependent factors and $\frac{b}{T^{2}}$ accounts for the temperature effect. There is another MTTF model for thin oxides:

\begin{equation}
MTTF_{OB} = A \exp(\frac{B}{E_{ox}}) \exp(\frac{Ea}{kT}),
\end{equation}

where $B$ stands for the electric field acceleration factor, $k$ for the Boltzmann's constant, and $T$ for the absolute temperature.

\subsection{Electromigration:}

The electromigration (EM) effect can play a critical role in circuit reliability at high current density  (typically exceeding 10$^{5}$ A/cm$%
^{2}$),\cite{ICCD1}. Energy and momentum from electrons can be transferred to metal atoms and lead to their biased movement towards the anode electrode. This mass transport through interface diffusion or grain boundary diffusion will create material depleted regions near the cathode and regions with access material near the anode. As a result, the layer could be damaged and the damage is manifested as open circuit failure near the cathode or short circuit failure near the anode regions. EM is a thermally activated process and the median time to failure (MTTF) can be described by the simplified formula

\begin{equation}
MTTF_{EM}=A*(j_e)^{-n}\exp (\frac{E_{a}}{kT})  
\end{equation}%
where $E_{a}$ is the activation energy of EM process, $j_e$ is the current density,  $A$ is a fabrication-dependent constant  and n can vary from $1 < n < 2$ depending on metal properties,\cite{ICCD1},\cite{ICCD5}.  Both the activation energy $E_{a}$ and $A$ constant depend on fabrication technology and are defined by material type, grain structure, size distribution, crystallographic orientation and impurities. The constant A also depends on the geometry and length of interconnects. It was shown, that activation energy can be reduced more than twice (from $1.4 eV - to - [0.5-0.8] eV$) by doping Al with a small amount of Cu (0.3-5\%),
\cite{ICCD1}.  In reality, the EM failure is a complicated process and non-Arrhenius temperature effects should be taken into consideration to predict the failure times. 

The poor metal quality, vacancies, and multiple irregular grain boundaries result in significant reduction of the activation energy that has severe effect on the median time to failure due to an exponential dependence. Electromigration induces stress gradients that compete with EM mass transfer by causing ion movement from compressive stress regions to regions with tensile stress. 

Material type and methods of interconnect deposition for submicron SMOC technologies (physical vapor deposition, PVD chemical vapor deposition, CVD; atomic layer deposition, ALD; electroplating; annealing; chemical-mechanical or electro-chemical-polishing) play critical role in grain microstructure of interconnects that determines the mass transport path and EM lifetimes.

\renewcommand\arraystretch{1.15}
\begin{table*}[htb]
\caption{Semiconductor Properties}
\label{TABDEGRADE1}
 \centering
 \begin{tabular}{ | p{1.00in} | p{2.00in} | p{1.50in} | p{1.50in} | } \hline
   \shortstack[l]{\\Degradation\\Mechanisms} & Lifetime Equation & Failure &
\shortstack[l]{\\Factors that effects\\critical parameters}\\ \hline
   Hot Carrier Injection (HCI)
   & \begin{equation*}
\small{MTTF_{HCI} = B(I_{sub}) - N \exp(\frac{E_a}{kT})}
\end{equation*}
 &  \begin{enumerate} \item \textbf{critical} device switching time delay 
 \item \textbf{non-responsive} device  functionality
 \end{enumerate}
 & \begin{enumerate}
\item Drain doping levels 
\item Channel lengths
\item Gate oxide thickness
\item Interface traps
\item Purity and quality of gate oxide 
\end{enumerate} 
 \\ \hline

   Oxide Breakdown (OB)
      &  \begin{equation*}
MTTF_{OB} = A \exp(\frac{B}{E_{ox}}) \exp(\frac{Ea}{kT})
\end{equation*}

 & \begin{enumerate} \item \textbf{short circuit} between the gate and the substrate \end{enumerate} & \begin{enumerate}
\item Purity of oxide layer Ð 
\begin{enumerate}
\item Crystal defects
\item Impurities (e.g., heavy metals) 
\item  Roughness of oxide surface 
\end{enumerate}
\item Gate oxide thickness
\end{enumerate}
\\ \hline

   Electromigration (EM)
      & \begin{equation*}
MTTF_{EM}=A*(j_e)^{-n}\exp (\frac{E_{a}}{kT})  
\end{equation*}
 & \begin{enumerate}
 \item \textbf{short circuit} conditions
 \item \textbf{open circuit} conditions
 \end{enumerate}
  & \begin{enumerate}
\item Interconnect metal type
\item Grain structure
\item Grain size distribution
\item Crystallographic orientation
\item Impurities
\item Geometry and length of interconnects
\end{enumerate}  \\ \hline

 Negative bias temperature instability  (NBTI)
      & \begin{equation*}
\Delta V_{th}=A(V_{g},t)\exp (-\frac{E_{NB}}{kT}) 
\end{equation*}
 & \begin{enumerate} \item \textbf{critical} device switching time delay 
 \item \textbf{non-responsive} device  functionality
 \end{enumerate}

 & \begin{enumerate}
\item  Nitrogen concentration near Si/SiO2 interface,
\item  Presence of boron
\item  Water near Si/SiO2 interface
\item   Gate dimensions
\item   Gate oxide thickness
    
\end{enumerate}  \\ \hline
 \end{tabular}
\end{table*}

\subsection{Negative bias temperature instability }

Negative Bias Temperature Instability (NBTI) is known since the very early developments of MOS devices; however, it is emerging as one of the major degradation mechanisms for deep-submicron CMOS technologies. NBTI causes a significant threshold voltage shift ($50-100 mV$) and decrease in drain current mostly in p-channel metal-oxide-semiconductor field effect transistors (pMOSFET) under a negative gate bias at elevated temperatures ($100-150^{\circ}{\rm C}$). The effect is due to creation of interface traps and to buildup of positive oxide charge over periods of time (from several months to years depending on the device operation conditions). Trapped holes can be thermally activated and can cause dissociation of oxide defects. The threshold voltage shift  $\Delta V_{th}$ is expressed as:

\begin{equation}
\Delta V_{th}=A(V_{g},t)\exp (-E_{NB}/kT)  \label{NB}
\end{equation}

where $E_{NB}$ is the NBTI activation energy and $A(V_{g},t)$ is a function of gate voltage and time,\cite{ICCD1},\cite{ICCD2},\cite{ICCD5}.

There are many models proposed to explain the NBTI effect, including oxide hole injection, electron tunneling and diffusion-reaction models.  The most accepted model is the diffusion-reaction (or electrochemical) concept that relates the activation energy $E_{NB}$  to diffusion of hydrogen species dissociate at the interface with multiple hydrogen-terminated $Si$ bonds ($Si-H$).  

Deposition processes for current MOSFETs employ oxynitride, $SiON$ layer as  a gate dielectric material deposited by plasma enhanced CVD or rapid thermal oxidation in presence of $NO$ or $NO_2$ gases. It was shown that while introduction of nitrogen into oxide layer improves transistor performance and increases dielectric constant of gate material, it also reduces NBTI lifetime by introducing bulk oxide defects and reducing thermal activation energy of defect generation at $Si/SiON$ interface. The threshold voltage shift and NBTI lifetime strongly depend on the interfacial nitrogen concentration at the $SiO_{2}/Si$ interface that can be controlled by temperature, pressure, duration of the thermal nitridation processing step. Typical range of the nitrogen concentration is around 2-15\%
 for the thermal nitridation process. Higher interfacial N concentrations are usually achieved using decoupled plasma nitridation. The higher nitrogen concentration, the larger  $\Delta V_{th}$ shift during NBTI stress.

\section{Critical Process Parameters}

In this section, the critical process factors, that yield maximum results for exploiting the wearing effects described in the previous sections, are investigated.  The exponential nature of lifetime prediction of the device for the wearing effects gives rise to Trojan alterations that greatly accelerate the wearing mechanics of the device.  

Table \ref{TABDEGRADE1} shows a list of factors that can be modified during the process manufacturing to induce reliability Trojans through the time-based wearing effects.  The table shows the degradation mechanism, the lifetime prediction model, the critical failures resulting from the degradation mechanism, and the critical factors.  

Knowing the process model, a process engineer can influence one or more steps in the IC fabrication to change one or more critical factors.  This change will inherently result in an accelerated MTTF model for one or more degradation mechanism.    The malicious process modification (reliability Trojan) will guarantee higher probability of the devices produced will have a greatly reduced lifetime, as shown in Fig. \ref{f1}.

\section {Future Work}

The process modifications of time based wear out mechanisms of HCI, NBTI, OB and EM will be simulated using TCAD to illustrate how variation in the IC process, effect device lifetimes. We will show how these wear out mechanism could be exploited for Trojans by design modifications. We will also introduce condition-based trigger class of Trojans, by investigating how the phenomena of ESD, Latchup and soft errors could be exploited as Trojans.  We will illustrate how packaging modifications could also be used to maliciously create Trojans.

\section{Conclusion}
This paper discussed the concept of a new class of hardware Trojan that exploits early wear out mechanisms in CMOS devices.  There are no detection techniques that can detect all the variations of these reliability based Trojans.  In any post-production tests the infected transistor are operating just as well as normal transistors with no discrepancy in performance.  The danger lies in the fact that until the Trojan transistor has been in operation for some time, the circuit appears as if no malicious alterations have been inserted.  These reliability Trojans are induced through minor variations in the manufacturing process.  Thus, there is a critical demand for detection techniques that properly identify these Trojans.

\bibliographystyle{IEEEtran}
\bibliography{iccd09}
\end{document}